\documentclass[lettersize,journal]{IEEEtran}
\usepackage{amsmath,amsfonts}
\usepackage{array}
\usepackage[caption=false,font=normalsize,labelfont=sf,textfont=sf]{subfig}
\usepackage{textcomp}
\usepackage{stfloats}
\usepackage{url}
\usepackage{verbatim}
\usepackage{graphicx}
\usepackage{cite}
\usepackage{multirow}
\usepackage{amssymb}
\usepackage{pifont}
\usepackage{longtable}
\usepackage{bbding}
\usepackage{tabularx}
\usepackage{tikz}
\usetikzlibrary{arrows.meta, positioning}
\usepackage{algpseudocode}
\usepackage{booktabs}

\usetikzlibrary{positioning, shapes.geometric}
\usepackage[linesnumbered,ruled,vlined]{algorithm2e}

\newcommand{\cmark}{\Checkmark}
\newcommand{\xmark}{\XSolid}
\newcommand{\cxmark}{\Checkmark\kern-1.2ex\raisebox{1ex}{\rotatebox[origin=c]{125}{\textbf{--}}}}

\begin{document}

\title{ AI Agent Access (A³) Network: An Embodied, Communication-Aware Multi-Agent Framework for 6G Coverage}

\author{Han Zeng, Haibo Wang,~\IEEEmembership{ Member,~IEEE,} Luhao Fan, Bingcheng Zhu, ~\IEEEmembership{Senior Member,~IEEE}, Xiaohu You$^{\ast}$,~\IEEEmembership{Fellow,~IEEE}, and Zaichen Zhang$^{\ast}$,~\IEEEmembership{Senior Member,~IEEE}
\thanks{This work is supported by the Fundamental Research Funds for the Central Universities 2242022k60001, NSFC project 62471126, Jiangsu Key R\&D Program Project BE2023011-2, National Key R\&D Program of China 2023YFB3609804, and the research fund of National Mobile Communications Research Lab. 2025A03. ($^{\ast}$Corresponding authors: Xiaohu You and Zaichen Zhang.)}

\thanks{Han Zeng, Haibo Wang, Bingcheng Zhu, Xiaohu You and Zaichen Zhang are with the National Mobile Communications Research Laboratory, Southeast University, Nanjing 210096, P.R.China, and they are also with the Purple Mountain Laboratory, Nanjing 211111, P. R. China. (e-mail: 230228224@seu.edu.cn, haibowang@seu.edu.cn, zbc@seu.edu.cn, xhyu@seu.edu.cn and zczhang@seu.edu.cn)
	
Luhao Fan is with the State Key Laboratory of Millimeter Waves, Southeast University, Nanjing 210096, P. R. China. (e-mail:lh.fan.ee@gmail.com)
}}

\markboth{Journal of \LaTeX\ Class Files,~Vol.~14, No.~8, August~2021}%
{Shell \MakeLowercase{\textit{et al.}}: A Sample Article Using IEEEtran.cls for IEEE Journals}

    \IEEEpubid{
	\begin{minipage}{\textwidth}
	    \centering
	    \footnotesize
	    \vspace{30pt} 
	    This work will be submitted to the IEEE for possible publication.  
	    Copyright may be transferred without notice, after which this version may no longer be accessible.
	\end{minipage}
	}

\maketitle

\begin{abstract}
	
	The vision of 6G communication demands autonomous and resilient networking in environments without fixed infrastructure. Yet most multi-agent reinforcement learning (MARL) approaches focus on isolated stages—exploration, relay formation, or access—under static deployments and centralized control, limiting adaptability. We propose the AI Agent Access (A³) Network, a unified, embodied intelligence-driven framework that transforms multi-agent networking into a dynamic, decentralized, and end-to-end system. Unlike prior schemes, the A³ Network integrates exploration, target user access, and backhaul maintenance within a single learning process, while supporting on-demand agent addition during runtime. Its decentralized policies ensure that even a single agent can operate independently with limited observations, while coordinated agents achieve scalable, communication-optimized coverage. By embedding link-level communication metrics into actor–critic learning, the A³ Network couples topology formation with robust decision-making. Numerical simulations demonstrate that the A³ Network not only balances exploration and communication efficiency but also delivers system-level adaptability absent in existing MARL frameworks, offering a new paradigm for 6G multi-agent networks.

\end{abstract}


\section{Introduction}

\IEEEPARstart{T}{he} envisioned Sixth Generation (6G) network aims to create a highly connected, intelligent, and resilient communication system that ensures seamless interaction among people, devices, and their surroundings~\cite{b0,b1}. Beyond speed and capacity, 6G prioritizes autonomy, real-time adaptability, and robust service in dynamic, uncertain conditions. In contrast, current wireless systems rely heavily on fixed infrastructure and static configurations. Even mobile elements like drones or vehicle relays typically follow preset paths or rule-based controls~\cite{b9,b11}. Such rigid, centralized designs lack the flexibility and intelligence to meet 6G’s evolving demands, underscoring the need for decentralized, self-organizing networks capable of autonomous, adaptive operation.

\begin{figure*}[!t]
	\centering
	\includegraphics[width=6in]{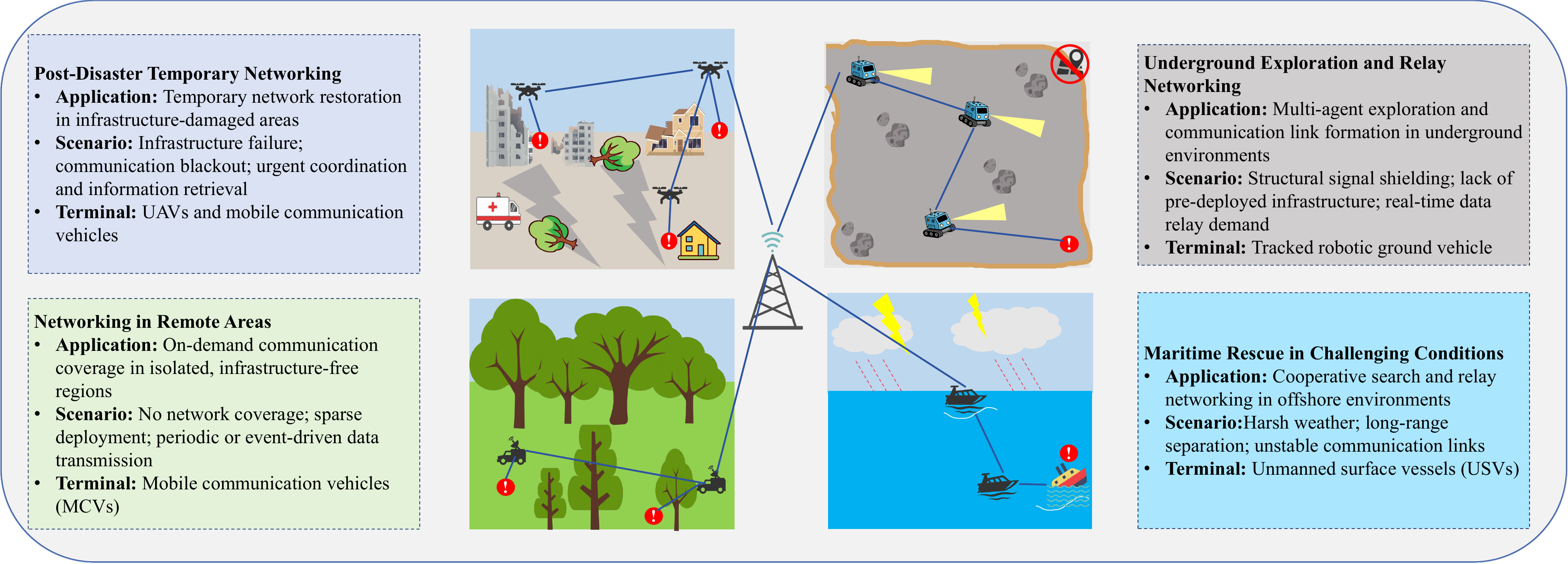}
	\caption{Background and Application scenarios.}
	\label{BG}
\end{figure*}

\begin{table*}
	\centering
	\scriptsize
	\caption{Summary of Multi-Agent Communication Schemes}
	\begin{tabularx}{\linewidth}{@{}l>{\arraybackslash}m{2.5cm}>{\centering\arraybackslash}m{0.5cm}>{\centering\arraybackslash}m{0.5cm}>{\centering\arraybackslash}m{0.5cm}>{\centering\arraybackslash}m{0.5cm}>{\arraybackslash}m{6.5cm}>{\arraybackslash}m{3.5cm}@{}}
		\hline
		Ref. & Network Topology & Comm. & Search & Deploy & Decent. & Methodology & Application \\
		\hline
		\cite{r1}  & Mesh                          & \cmark & \xmark & \cxmark & \cxmark & Multi-Agent Actor-Critic Reinforcement Learning (MACRL) & Dynamic AP optimization \\
		\cite{r2}  & Ad-hoc                        & \cmark & \xmark & \cxmark & \cmark  & Multi-Agent Reinforcement Learning (MARL), Q-learning & UAV-user association \\
		\cite{r4}  & Ad-hoc                        & \cmark & \xmark & \cmark & \cxmark & Federated Soft Actor-Critic (FeSAC) & Train-to-train communication \\
		\cite{r5}  & Ad-hoc                        & \xmark & \cmark & \cmark & \cmark  & Distributed Deep Q-Learning (DDQL) with shared memory & UAV swarm surveillance \\
		\cite{r6}  & Ad-hoc                        & \cmark & \xmark & \cmark & \cxmark  & K-Means-enhanced Multi-Agent PPO (KMAPPO) & Emergency communication \\
		\cite{r7}  & Ad-hoc                        & \cmark & \xmark & \cmark & \cmark  & Multi-Agent QMIX (MAQMIX) with inter/intra training & UAV packet routing \\
		\cite{r10} & Heterogeneous Network         & \cmark & \xmark & \cxmark & \cxmark & QMIX & Satellite beam hopping \\
		\cite{r11} & Hybrid Relay Topology         & \cmark & \xmark & \cmark & \cxmark & Multi-Agent Deep Deterministic Policy Gradient (MADDPG), Proximal Policy Optimization (PPO) & Maritime optical communication \\
		\cite{r12} & Ad-hoc                        & \cmark & \xmark & \cmark & \cmark  & Deep Deterministic Policy Gradient (DDPG) & UAV coverage \\
		\cite{r13} & Coop. Network                 & \cmark & \xmark & \cmark & \cxmark & Q-learning, Actor-Critic & UAV-assisted computing \\
		\cite{r14} & Heterogeneous Network         & \cmark & \xmark & \cxmark & \xmark  & MADDPG & Satellite-terrestrial NOMA \\
		\cite{r15} & Centralized Topology & \cmark & \xmark & \cxmark & \cxmark & Deep Q-Network (DQN) & RAN slicing \\
		\cite{r16} & Coop. Network                 & \cmark & \cmark & \cmark & \cxmark & Multi-Agent Twin Delayed DDPG (TD3) & UAV task offloading \\
		\cite{r17} & Mesh                          & \xmark & \xmark & \cxmark & \cmark  & Particle Swarm Optimization (PSO) with adaptive learning & Distributed optimization \\
		\cite{r18} & Mesh                          & \cmark & \xmark & \cxmark & \cmark  & DQN & Dense cell networks \\
		\cite{r19} & Mesh                          & \cmark & \xmark & \cxmark & \cxmark & MARL, PPO & UAV cellular networks \\
		\cite{r21} & Mesh                          & \xmark & \cmark & \cmark & \cxmark & Asynchronous Advantage Actor-Critic (A3C), CNN & Federated UAV learning \\
		\cite{r22} & Heterogeneous Network         & \xmark & \cmark & \cmark & \cxmark & MA2DDPG, AC-Mix & UAV-edge computing \\
		\cite{r23} & Coop. Network                 & \cmark & \xmark & \cmark & \cxmark  & Q-learning, Actor-Critic & UAV trajectory planning \\
		\cite{r24} & Coop. Network                 & \cmark & \xmark & \cxmark & \cxmark & DQN, Advantage Actor-Critic (A2C) & Dense cell scheduling \\
		\cite{r25} & Coop. Network                 & \cmark & \cmark & \cxmark & \cxmark & Multi-Agent Soft Actor-Critic (MASAC) & IIoT federated learning \\
		\cite{r27} & Heterogeneous Network         & \cmark & \xmark & \cxmark & \cxmark & MDRL, DQN-OS, DDPG-OS & Resource allocation \\
		\cite{r28} & Hybrid Hierarchical           & \cmark & \xmark & \cmark & \cxmark & MADDPG & Disaster rescue with UAV \\
		\hline
		Ours       & Hybrid Hierarchical           & \cmark & \cmark & \cmark & \cxmark & Embodied MARL, Actor-Critic & Generic agent access \\
		\hline
	\end{tabularx}
	\label{tab:multiagent-summary}
	
	\vspace{-0.5em}
	\begin{flushleft}
		\footnotesize
		\textbf{Legend:} \\
		\textbf{Comm.} — Whether communication performance is explicitly optimized (vs.\ resource allocation or other metrics). \\
		\textbf{Search} — Whether the method requires agents to explore unknown environments or targets (vs.\ known, static targets). \\
		\textbf{Deploy} — Agent deployment flexibility; \cxmark{} means parameters tunable but fixed agent count/location, \cmark{} means dynamic repositioning and reconfiguration; only our method supports dynamic addition of agents during runtime. \\
		\textbf{Decent.} — Decentralization level; \cxmark{} indicates partial decentralization (e.g., CTDE). \\
	\end{flushleft}
\end{table*}

Building on the limitations of traditional wireless systems discussed above, AI-driven multi-agent networks offer a promising path forward. By enabling agents to perceive their environments, make decisions in real time, and collaborate autonomously, these networks support robust communication in complex, dynamic scenarios~\cite{b3,b4}. Using multi-agent reinforcement learning (MARL)\cite{b2,b6,b7}, they have been applied to rapid post-disaster deployment, underground relay in GPS-denied areas, remote connectivity, and cellular network enhancement (see Fig.\ref{BG}). Existing works (Table~\ref{tab:multiagent-summary}) cover diverse network types—including ad-hoc~\cite{r2,r4,r5,r7,r12}, mesh~\cite{r1,r17,r18,r21}, cooperative~\cite{r13,r16,r23,r24,r25}, and heterogeneous networks~\cite{r10,r14,r22}—and employ various MARL algorithms, such as value-based (DQN, QMIX~\cite{DRL,QMIX}) and actor-critic methods (PPO, MADDPG~\cite{PPO,MADDPG}). Among them, actor-critic approaches are especially effective for decentralized learning under partial observability.

Despite rapid progress in MARL-based wireless control, existing methods face fundamental barriers that prevent their deployment in dynamic, infrastructure-free 6G environments. Most prior works rely on static or pre-assigned agent placements and often assume prior knowledge of user locations, which precludes autonomous exploration and on-demand resource allocation in unknown or rapidly evolving settings~\cite{r17,r18,r19,r24,r25}. Furthermore, essential processes such as relay formation, exploration, and agent deployment are typically studied in isolation rather than as components of a coherent system, resulting in fragmented solutions with poor cross-stage coordination. The absence of communication-aware learning further limits generalization, confining policies to fixed team sizes or pre-defined topologies and severely undermining robustness and scalability.

Embodied intelligence emphasizes the tight coupling of perception, action, and environmental feedback, providing a foundation for adaptive multi-agent networks. Leveraging this principle allows agents to respond to environmental cues in real time, coordinate across previously isolated stages, and support dynamic, on-demand deployment.

In our proposed \textit{A³ Network} (AI Agent Access Network), this principle guides the design of a unified, perception-driven, embodied multi-agent framework that integrates environment-aware exploration, dynamic agent deployment, and adaptive topology formation into a single end-to-end learning process. While embodied intelligence informs agent-level adaptability, the overall system performance emerges from the combination of decentralized policies, hybrid hierarchical topology, and communication-aware reinforcement learning, collectively overcoming the shortcomings of existing MARL frameworks. Key innovations include:  

\textbf{(1) Embodied, perception-driven agent deployment:} Unlike existing methods that rely on fixed deployments or pre-assigned agent roles, our agents actively sense and interpret their radio environment to autonomously explore unknown areas and trigger on-demand deployment of additional agents. This real-time, environment-responsive strategy enables efficient resource use and flexible coverage extension, moving beyond static or rule-based approaches to achieve truly adaptive, scalable network formation.
	
\textbf{(2) Hybrid hierarchical topology with structural–functional separation:} We introduce a hybrid hierarchical topology that separates the \textit{Communication Graph} (physical connectivity and link states) from the \textit{Logical Control Tree} (task allocation and coordination). This separation enables topology-aware optimization without entangling it with task scheduling, allowing the network to scale efficiently and integrate new agents seamlessly. Such structure–function decoupling is particularly valuable for communication systems facing rapidly changing topologies and heterogeneous mission demands.
	
\textbf{(3) Communication-aware reinforcement learning with structural generalization:} We develop a communication-aware MARL framework where GNN-based actor–critic models are trained under centralized supervision to process time-varying network graphs annotated with link metrics (capacity, latency, load). The critic evaluates global utility over the dynamic topology, while the actor aggregates local neighborhood features to make routing and power-control decisions. This structure facilitates transferring policies trained on smaller networks to larger, heterogeneous topologies with minimal adaptation, mitigating a key scalability challenge faced by prior RL-based wireless control schemes.

The rest of the paper is organized as follows. Section~II describes the system model and problem formulation. Section~III introduces the decentralized decision-making mechanism based on an actor-critic architecture within the A³ network. Section~IV presents the overall system workflow, including training and testing procedures. Section~V reports the numerical results, and Section~VI concludes the paper.

\section{System Model and Problem Formulation}

\begin{figure*}[!t]
	\centering
	\includegraphics[width=6in]{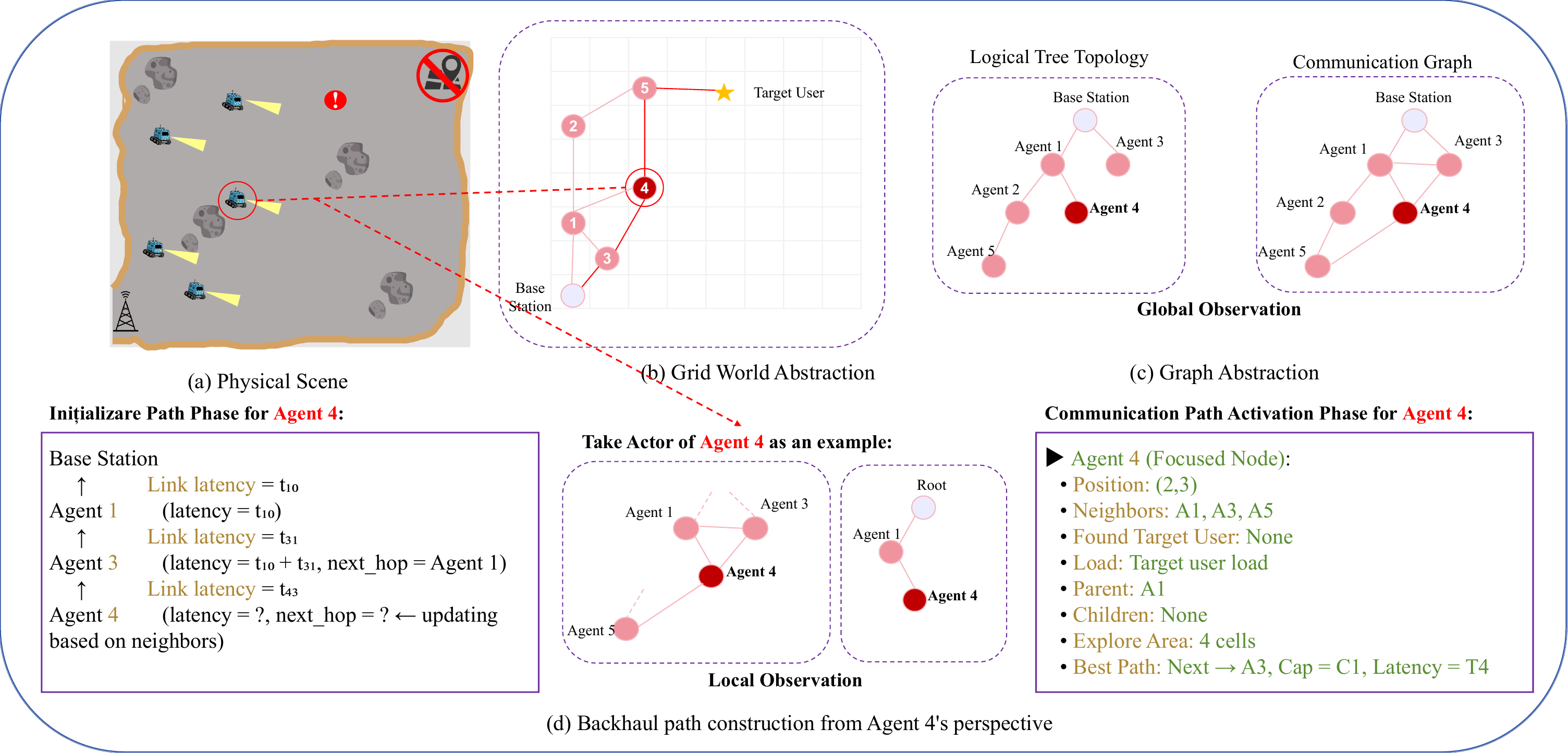}
	\caption{System Abstraction and Path Management Overview.}
	\label{Gh}
\end{figure*}

The proposed system enables unified exploration, access, and networking by deploying intelligent agents that autonomously operate in unknown environments without prior information about target user locations. To formalize this design, in this paper, we make the following assumptions:

\begin{enumerate}
	\item \textbf{Grid-Based Environment}: The space is discretized into grid cells, with agents and target users occupying distinct positions. Agent movement is restricted to the five primitive actions $\{\text{up}, \text{down}, \text{left}, \text{right}, \text{stay}\}$, each corresponding to one grid step. Manhattan distance is used to characterize perception range and feasible reachability under this grid-based motion model, while Euclidean distance is adopted for communication-related metrics such as signal strength and path loss.
	
	\item \textbf{Agent Model}: Agents are homogeneous, mobile, and equipped with sensing and communication modules. They operate under the decentralized partially observable Markov decision process (Dec-POMDP) settings without prior knowledge of target users. Communication links are interference-free within radius $r_c$. New agents can be dynamically deployed by the base station.
	
	\item \textbf{Local Decision and Safe Execution}: Each agent selects actions independently based on local observations. Actions are executed sequentially, with constraint checks. Invalid actions (e.g., those breaking connectivity) are ignored, and the agent retains its previous state.
\end{enumerate}

The system consists of three modules: environment, communication, and agent. The environment setup defines spatial structure and boundary constraints; the communication model captures information exchange under physical limits; the agent behavior module describes perception, action, and coordination. 

\subsection{Environment Setup}

As illustrated in Fig.~\ref{Gh}(a), intelligent vehicles equipped with sensing and communication capabilities act as agents to explore signal‑deprived underground environments and maintain a link to the base station throughout the mission. The physical space is then discretized into a grid‑based environment, as shown in Fig.~\ref{Gh}(b).

The environment is defined as a two-dimensional discrete grid \(\mathcal{E} \subset \mathbb{Z}^2\), where each cell may be occupied by an agent, a target user, or both. Users are randomly distributed and represented by \(\mathcal{T} = \{\tau_1, \ldots, \tau_{|\mathcal{T}|}\}\), with each target user \(\tau_i\) characterized by its position \(\mathbf{p}_{\tau_i} \in \mathcal{E}\), workload \(L_i\), minimum channel requirement \(C_i^{\min}\), latency bound \(D_i^{\max}\), and priority level \(P_i\).

The system is initialized from a base station located at \(\mathbf{p}_\mathrm{BS} = (0, 0)\). Agents are deployed dynamically during exploration as needed, forming a time-varying set \(\mathcal{A}(t) = \{a_1, \ldots, a_{N(t)}\}\), where each agent \(a_i\) occupies position \(\mathbf{p}_i(t) \in \mathcal{E}\). The team size \(N(t)\) evolves over time based on the current exploration status and task demands.

\subsection{Communication Model} 

We adopt a hybrid topology consisting of two coupled graph structures: a dynamic \textit{physical communication graph} \(\mathcal{G}_c(t)\) and a \textit{logical control tree} \(\mathcal{T}_c(t)\), as illustrated in Fig.~\ref{Gh}(c). Each captures different aspects of the network, together supporting decentralized coordination and adaptive task allocation.

\paragraph{Logical Control Tree \(\mathcal{T}_c(t)\)}
Rooted at the base station, the logical tree \(\mathcal{T}_c(t)\) hierarchically organizes agents to enable decentralized role assignment and task propagation. New agents are incrementally added as leaf nodes based on exploration demands. Maintained independently from physical connectivity, the tree ensures each parent-child pair has a reliable communication link for effective coordination. Its structured, goal-oriented branching promotes efficient coverage and coordinated expansion into unexplored or disconnected regions.

\paragraph{Physical Communication Graph \(\mathcal{G}_c(t)\)}

Each agent \(a_i \in \mathcal{A}(t)\) establishes communication links with neighboring agents located within a communication radius \(r_c\). An undirected edge \((a_i, a_j) \in \mathcal{E}_c(t)\) is formed if \(\|\mathbf{p}_i(t) - \mathbf{p}_j(t)\|_1 \leq r_c\) (using Manhattan distance over the discrete grid), resulting in the dynamic physical communication graph \(\mathcal{G}_c(t) = (\mathcal{A}(t), \mathcal{E}_c(t))\), which reflects the instantaneous wireless connectivity among agents.

\paragraph{Relay Path Construction} 

Each wireless link \((a_i, a_j) \in \mathcal{E}_c(t)\) is modeled as an additive white Gaussian noise (AWGN) channel. The received signal at agent \(a_j\) from agent \(a_i\) is given by:
\begin{equation}
	y_j = h_{ij} x_i + n_j,
\end{equation}
\noindent where \(h_{ij}\) denotes the distance-dependent channel gain, and \(n_j \sim \mathcal{N}(0, \sigma^2)\) is Gaussian noise.

Under the path loss model, the instantaneous signal-to-noise ratio (SNR) of the link is:
\begin{equation}
	\mathrm{SNR}_{ij}(t) = \frac{P_i G_t G_r \lambda^2}{(4\pi)^2 \sigma^2 \|\mathbf{p}_i(t) - \mathbf{p}_j(t)\|_2^2},
\end{equation}
\noindent capturing the combined effects of transmit power, antenna gains, wavelength, path loss, and noise. The corresponding theoretical link capacity is:
\begin{equation}
	C_{ij}(t) = B \log_2 \left(1 + \mathrm{SNR}_{ij}(t)\right).
\end{equation}

In practice, the effective capacity available for relaying data depends on current load conditions. Denote by \(L_{ij}(t)\) the load occupying link \((a_i, a_j)\), then the available capacity is:
\begin{equation}
	C_{ij}^{\mathrm{avail}}(t) = C_{ij}(t) - L_{ij}(t).
\end{equation}

The end-to-end delay, combining propagation and transmission delays, is:
\begin{equation}
	D_{\tau \rightarrow \mathrm{BS}}(t) = \sum_{(a_i, a_j) \in \mathcal{P}_{\tau \rightarrow \mathrm{BS}}} \left( \frac{d_{ij}(t)}{v} + \frac{L_i}{C_{ij}^{\mathrm{avail}}(t)} \right),
\end{equation}
\noindent where \(d_{ij}(t) = \|\mathbf{p}_i(t) - \mathbf{p}_j(t)\|_2\) is hop distance, \(v\) is propagation speed, and \(L_i\) is the data load for the target user \(\tau_i\).

The bottleneck capacity along the path is then the minimal available capacity:
\begin{equation}
	C_{\tau \rightarrow \mathrm{BS}}^{\mathrm{avail}}(t) = \min_{(a_i, a_j) \in \mathcal{P}_{\tau \rightarrow \mathrm{BS}}} C_{ij}^{\mathrm{avail}}(t).
\end{equation}

To ensure relay feasibility, the load must not exceed any relay's capacity:
\begin{equation}
	L_i \leq L_k^{\max}(t), \quad \forall a_k \in \mathcal{P}_{\tau \rightarrow \mathrm{BS}},
\end{equation}
where \(L_k^{\max}(t)\) denotes per-agent capacity limits.

Finally, each agent must maintain a valid path to the base station to guarantee network connectivity:
\begin{equation}
	\forall a_i \in \mathcal{A}(t), \quad \exists\, \mathcal{P}_{i \rightarrow \mathrm{BS}}(t) \subseteq \mathcal{G}_c(t).
\end{equation}

\subsection{Agent Behavior and Decision Process}

We model agent behavior as Dec-POMDP, which captures local observability, uncertainty, and decentralized decision-making. Formally, the Dec-POMDP is defined as:
\begin{equation}
	\left( \mathcal{S}, \{\mathcal{O}_i\}_{i=1}^N, \{\mathcal{U}_i\}_{i=1}^N, P, R, \gamma \right),
\end{equation}
\noindent where:
\begin{itemize}
	\item \(\mathcal{S}\): global state space, including the complete environment configuration—agent positions, latent target user locations, link connectivity, and environmental features. Although user positions are part of \(\mathcal{S}\), they are initially unknown to agents and must be discovered through exploration;
	\item \(\mathcal{O}_i\): local observation space of agent \(a_i\), determined by its sensing and communication range, and derived as a partial projection of \(\mathcal{S}\);
	\item \(\mathcal{U}_i = \mathcal{U}_i^{\mathrm{move}} \times \mathcal{U}_i^{\mathrm{request}}\): the agent's joint action space, comprising a discrete movement action ($\mathcal{U}_i^{\mathrm{move}} = \{\text{up}, \text{down}, \text{left}, \text{right}, \text{stay}\}$) and a binary request decision indicating whether to initiate a new agent deployment request;
	\item \(P(s' \mid s, \mathbf{u})\): transition function that determines the next state based on current state \(s\) and joint action \(\mathbf{u} = (u_1, \ldots, u_N)\);
	\item \(R(s, \mathbf{u})\): global reward function shared across agents, reflecting system performance in terms of coverage, connectivity, and deployment efficiency;
	\item \(\gamma \in (0,1]\): discount factor for future rewards.
\end{itemize}

The system is in state \(s_t\), each agent \(a_i\) receives a local observation \(o_i^t \in \mathcal{O}_i\) and selects an action \(u_i^t = (u_i^{\mathrm{move}}, u_i^{\mathrm{dispatch}})\). The joint action \(\mathbf{u}_t = (u_1^t, \ldots, u_N^t)\) induces a state transition \(s_{t+1} \sim P(\cdot \mid s_t, \mathbf{u}_t)\), yielding a global reward \(r_t = R(s_t, \mathbf{u}_t)\).

\subsection{Problem Formulation}

Let \(\mathcal{T} = \{\tau_1, \ldots, \tau_{|\mathcal{T}|}\}\) be the target users and \(\mathcal{P}_{\tau_i \rightarrow \mathrm{BS}}\) their relay paths to the base station. The objective is to maximize the minimum link capacity across all paths while ensuring access and communication constraints:

\begin{equation}
	\begin{aligned}
		&\max_{\{\mathcal{P}_{\tau_i \rightarrow \mathrm{BS}}\}} \quad 
		\min_{\tau_i \in \mathcal{T}} \left( \min_{(a_k, a_{k+1}) \in \mathcal{P}_{\tau_i \rightarrow \mathrm{BS}}} C_{k,k+1}^{\mathrm{avail}}(t) \right) \\
		\text{s.t.} \quad 
		& \left\{
		\begin{alignedat}{2}
			& \frac{|\mathcal{T}_{\text{connected}}(t)|}{|\mathcal{T}|} \geq \rho_{\min}              && \hspace{0 em} \text{(C1)} \\
			& D_{\tau_i \rightarrow \mathrm{BS}}(t) \leq D_i^{\max}, \quad \forall \tau_i \in \mathcal{T} && \text{(C2)} \\
			& C_{k,k+1}^{\mathrm{avail}}(t) \geq C_i^{\min}, \quad \forall (a_k, a_{k+1}) \in \mathcal{P}_{\tau_i \rightarrow \mathrm{BS}} && \text{(C3)} \\
			& L_i \leq L_k^{\max}(t), \quad \forall a_k \in \mathcal{P}_{\tau_i \rightarrow \mathrm{BS}} && \text{(C4)} \\
			& \|\mathbf{p}_k(t) - \mathbf{p}_{k+1}(t)\|_1 \leq r_c, \quad \forall (a_k, a_{k+1}) \in \mathcal{P}_{\tau_i \rightarrow \mathrm{BS}} && \text{(C5)} \\
			& \exists\, \mathcal{P}_{a_j \rightarrow \mathrm{BS}} \subseteq \mathcal{G}_c(t), \quad \forall a_j \in \mathcal{A}(t) && \text{(C6)} \\
			& \mathbf{p}_j(t) \in \mathcal{E}, \quad \forall a_j \in \mathcal{A}(t) \cup \mathcal{T} && \text{(C7)}
		\end{alignedat}
		\right.
	\end{aligned}
\end{equation}

\noindent where \(\mathcal{T}_{\text{connected}}(t)\) denotes the set of target users discovered and connected to the base station at time \(t\). Constraints (C1)–(C7) ensure system feasibility: (C1) enforces a minimum target user access rate for effective exploration; (C2)–(C4) regulate delay, available link capacity, and load on relay paths; (C5)–(C7) limit communication range, guarantee global connectivity, and ensure valid deployment. These conditions jointly capture exploration, communication, and physical constraints, but assume full knowledge of target users and relay paths. In practice, they mainly serve as a benchmark to guide reward design and performance evaluation.

\section{Decentralized Decision Making via Actor-Critic Learning in A³ Network}

\begin{figure*}[!t]
	\centering
	\includegraphics[width=5.5in]{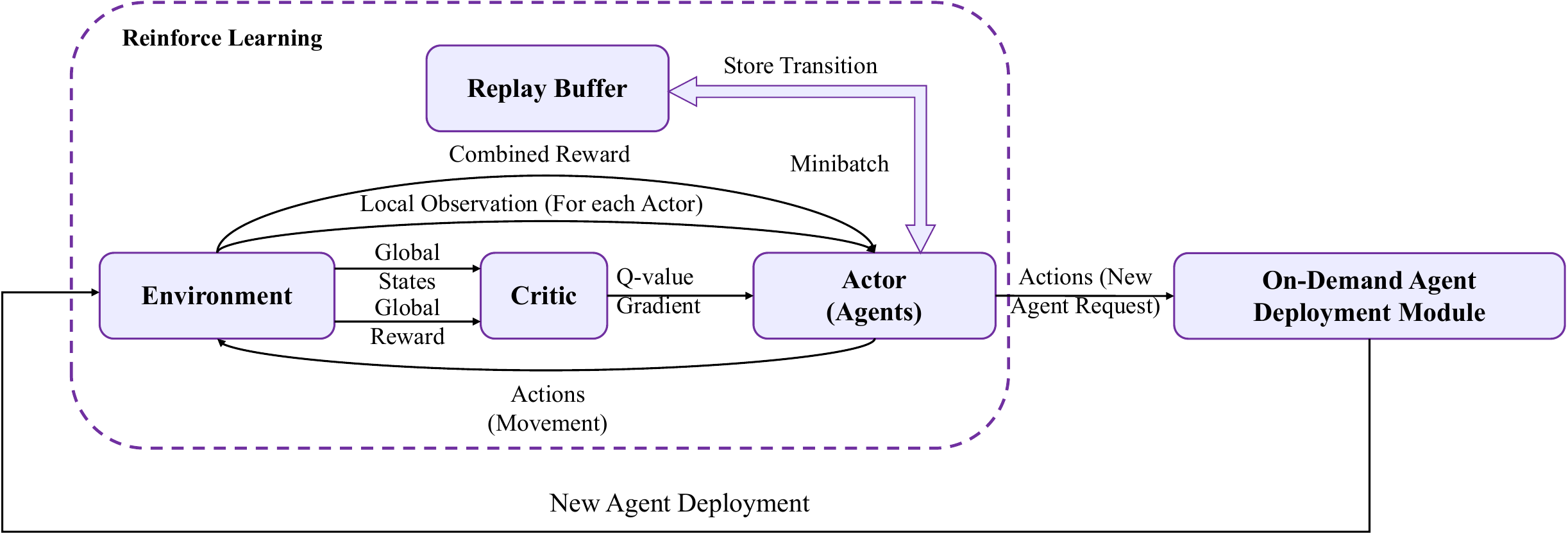}
	\caption{Actor-Critic Architecture Design.}
	\label{AC}
\end{figure*}

\begin{figure}[!t]
	\centering
	\includegraphics[width=3.2in]{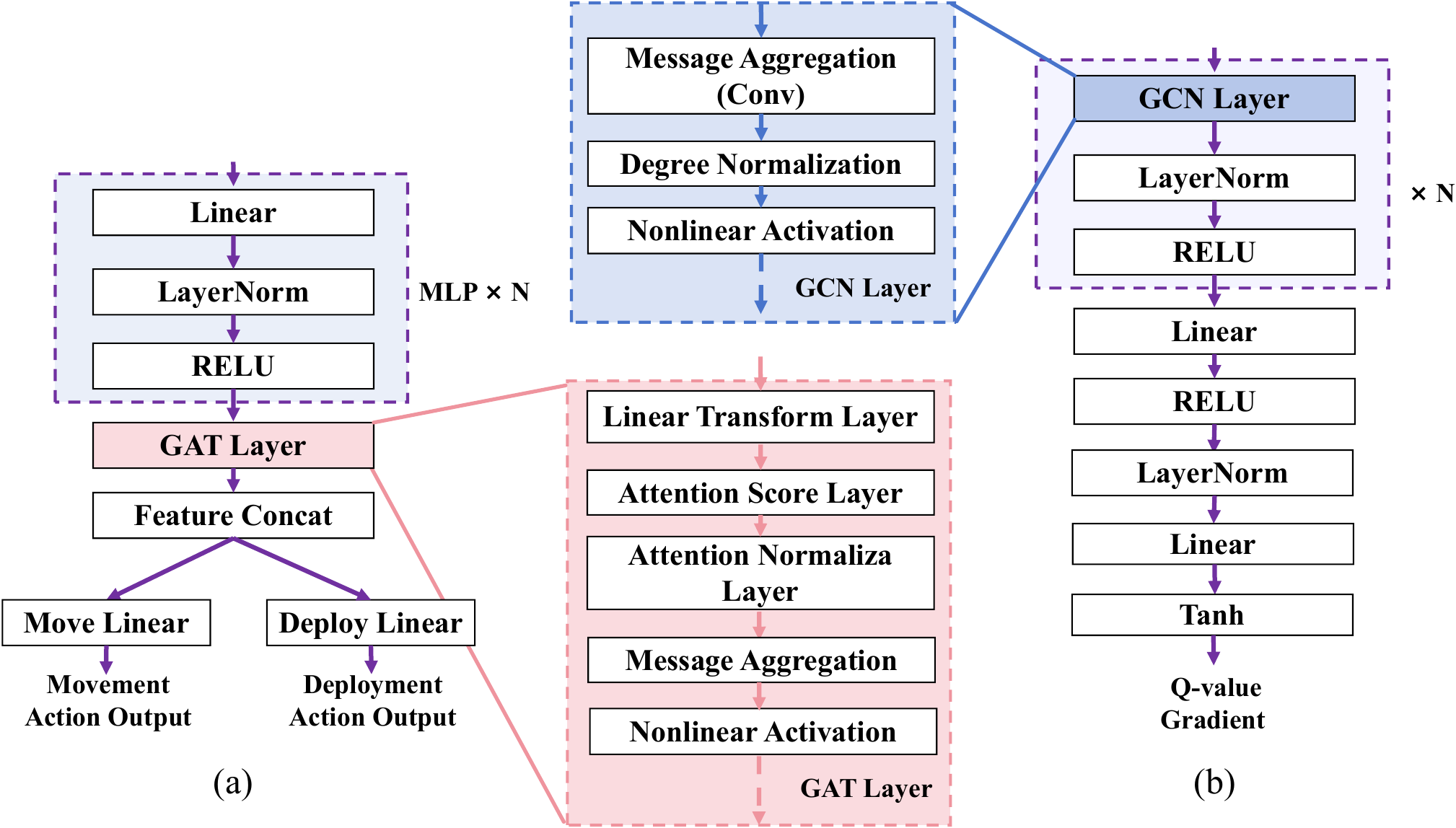}
	\caption{Network Architectures of Actor and Critic in the A³ Network (a) Structure of the \textbf{actor} network; (b) Structure of the \textbf{critic} network.}
	\label{NW}
\end{figure}

As illustrated in Fig.~\ref{AC}, the proposed \textbf{A³ Network Structure} adopts an actor-critic architecture under the \textbf{Centralized Training with Decentralized Execution} (CTDE) paradigm. It consists of four main modules: Environment, Actor, Critic, and Replay Buffer. Agents make decisions based on local observations, including both movement and deployment requests that trigger an on-demand dispatch mechanism for dynamically adding new agents. The system incorporates structural awareness and communication feedback, enabling scalable coordination under partial observability. 

\subsection{Actor-Critic Architecture with GNN-based Design} 

Graph neural networks (GNNs) are integrated into both the actor and the critic to effectively handle dynamically changing agent populations and input sizes. Unlike traditional networks that require fixed input dimensions, GNNs naturally adapt to variable graph structures by aggregating neighborhood information, enabling scalable and flexible policy learning in dynamic environments~\cite{b5,b10}.

Fig.~\ref{NW} illustrates the actor and critic network architectures used in our CTDE framework. Subfigure (a) depicts the actor’s decentralized policy network, while subfigure (b) depicts the centralized critic network leveraging global information. 

\paragraph{Critic Design}

The centralized critic employs GNN layers to capture inter-agent interactions and relational information at each time step \(t\). Specifically, each agent \(i\) in the active agent set \(\mathcal{A}(t)\) generates an embedding vector \(h_i^t\), which aggregates features from its neighbors according to the communication graph topology. These embeddings encode both the agent's local state and its relational context.

To produce a scalar Q-value, each embedding is first transformed by a shared learnable mapping followed by a nonlinear activation \(\sigma\), and then aggregated across all active agents using a permutation-invariant operator (mean pooling):
\begin{equation}
	Q_\phi(s, \mathbf{u}) = \mathrm{MEAN}\!\left( \left\{ \sigma(W_h h_i^t + b_h) \right\}_{i \in \mathcal{A}(t)} \right),
\end{equation}
\noindent where \(W_h\) and \(b_h\) are learnable parameters, and \(\sigma\) denotes an activation function such as ReLU. The pooling step ensures that the critic remains robust to varying team sizes and agent orderings. 

Here, \(h_i^t\) is the embedding vector of agent \(i\) at time \(t\), obtained by aggregating its local observation–action pair \((o_i^t,u_i^t)\) and neighborhood features through the GNN layers. Thus, the dependence on the global state \(s\) and joint action \(\mathbf{u}\) is implicitly captured in the set of embeddings \(\{h_i^t\}_{i \in \mathcal{A}(t)}\).

The temporal difference (TD) target for training is computed as:
\begin{equation}
	y_k = r_k + \gamma Q_{\phi'}\!\big(s_k', \{\mu_{\theta_j}(o_j^{\,\prime})\}_{j \in \mathcal{A}(t+1)}\big),
\end{equation}
\noindent where \(r_k\) is the reward at step \(k\), \(\gamma \in (0,1)\) is the discount factor, \(\phi'\) are the parameters of a target network (a periodically updated copy of \(\phi\)) for stabilizing training, \(s_k'\) is the next global state, and \(\mu(s_k')=\{\mu_{\theta_j}(o_j^{\,\prime})\}_{j \in \mathcal{A}(t+1)}\) denotes the joint action produced by all target actors.

The critic parameters \(\phi\) are optimized by minimizing the mean squared TD error over a batch of size \(B_{\mathrm{batch}}\):
\begin{equation}
	\mathcal{L}_{\text{critic}} = \frac{1}{B_{\mathrm{batch}}} \sum_{k=1}^{B_{\mathrm{batch}}} \left(y_k - Q_\phi(s_k, \mathbf{u}_k)\right)^2,
\end{equation}
where \((s_k, \mathbf{u}_k)\) are sampled from the centralized replay buffer.

The parameters \(\phi\) are updated via gradient descent:
\begin{equation}
	\phi \leftarrow \phi - \eta \nabla_\phi \mathcal{L}_{\text{critic}},
\end{equation}
where \(\eta > 0\) is the critic learning rate.

\paragraph{Actor Design}

Each agent \(a_i\) maintains a decentralized stochastic policy \(\pi_{\theta_i}\) that maps its local observation \(o_i^t\) to a compound action \(u_i^t = (\Delta_i^t, r_i^t)\), where \(\Delta_i^t\) denotes a discrete movement direction, and \(d_i^t \in \{0,1\}\) is a binary deployment request.

Since the number of communication neighbors \(|\mathcal{N}_i^t|\) varies dynamically, the actor also incorporates a GNN to process neighborhood information in a scalable manner. Each local observation is first encoded through a shared linear transformation, followed by feature aggregation using a graph attention layer. The resulting embedding is concatenated with the agent’s own features and fed into two task-specific heads: one for movement and the other for deployment. Formally:
\begin{equation}
	u_i^t = \mu_{\theta_i}(o_i^t) = \left[ \mathrm{OneHot}(\Delta_i^t) \,\|\, d_i^t \right].
\end{equation}

To address training instability caused by partial observability and non-stationarity, all agents share a centralized replay buffer \(\mathcal{D}\). Each entry in \(\mathcal{D}\) is stored as:  
\[
(s, \mathbf{o}, \mathbf{u}, r, s', \mathbf{o}'),
\]  
\noindent where \(s\) and \(s'\) are the global state and next state, \(\mathbf{o} = (o_1, \ldots, o_N)\) and \(\mathbf{o}'\) are the joint local observations, \(\mathbf{u} = (u_1, \ldots, u_N)\) is the joint action, and \(r\) is the global reward. During training, the centralized critic leverages \((s, \mathbf{u}, r, s')\) for value estimation, while each actor is optimized only with its own local observation–action pairs \((o_i, u_i)\).

Although each agent follows a stochastic policy during exploration, we employ a deterministic policy gradient update for efficiency and stability, following the standard CTDE practice. The gradient for agent \(i\) is computed as:
\begin{equation}
	\nabla_{\theta_i} J(\theta_i) = \frac{1}{B_{\mathrm{batch}}} \sum_{k=1}^{B_{\mathrm{batch}}} 
	\nabla_{u_i} Q(s_k, \mathbf{u}) \Big|_{\mathbf{u}=\mu(\mathbf{o}_k)} 
	\cdot \nabla_{\theta_i} \mu_{\theta_i}(o_i^k),
\end{equation}
and the parameters \(\theta_i\) are updated via:
\begin{equation}
	\theta_i \leftarrow \theta_i + \alpha \nabla_{\theta_i} J(\theta_i),
\end{equation}
where \(\alpha > 0\) is the actor learning rate.

\paragraph{Centralized Training with Decentralized Execution}

Our framework follows the CTDE paradigm: a single centralized critic, shared across all agents, leverages full global information during training, including the complete communication graph, joint states, and actions. In contrast, each agent maintains an independent actor that conditions only on local observations and messages from neighbors during execution. This separation enables globally informed value estimation while preserving decentralized decision-making. Sharing a unified critic across agents improves sample efficiency, learning stability, and scalability, while avoiding retraining when the team size or topology changes.

\subsection{Communication Indicator and Reward Design} 

The formal objective in Section~II maximizes the minimum bottleneck capacity across target user-to-base paths, but it assumes that all target users have been discovered and their relay paths fully established—an assumption that rarely holds during early exploration. To address this, we design the reward structure as a principled surrogate: the \textbf{global reward} softly approximates key objective terms such as connectivity, target user access, and load balance, while \textbf{local role-based rewards} guide agent behaviors like frontier expansion and relay stabilization. Although the final objective is not directly computable during deployment, incremental, role-consistent rewards provide real-time feedback that asymptotically aligns policy learning with long-term system goals.

\paragraph{Role-Based Local Reward Structuring}

Agents dynamically assume one of two roles—\emph{explorer} or \emph{relay}—based on their relative depth in the communication tree topology. Leaf-level agents serve as explorers, focusing on unknown space exploration and target user discovery. Upper-layer agents function as relays to preserve hierarchical connectivity and ensure communication robustness. An explorer transitions into a relay when it successfully deploys a new agent as its direct child, thereby establishing a local subtree structure. Each agent \(i\)'s local reward at time \(t\), denoted \(r_i^t\), is computed conditionally on its role as explorer reward $r^{\text{explorer}}_i(t)$ or relay reward $r^{\text{relay}}_i(t)$.

\paragraph*{\textbf{Explorer Reward}}

For explorer agents, the reward function is designed to incentivize efficient frontier expansion while discouraging redundant movement and excessive clustering. The primary objective is to guide exploration toward sparsely populated or unvisited regions, while also promoting the timely discovery of valuable target users. The local reward for an explorer agent $i$ at time $t$ is defined as:

\begin{equation}
	r^{\text{explorer}}_i(t) = w_1 \cdot \Delta E_i^t + w_2 \cdot \mathbb{I}_{\{\text{target user found}\}} - w_3 \cdot \rho_i^t - \mathcal{P}_i^t,
\end{equation}

\noindent where \(\Delta E_i^t\) is the number of newly explored cells at time \(t\), \(\mathbb{I}_{\{\text{target user found}\}}\) stands for the indicator function that returns 1 if a new target user is found by agent \(i\), \(\rho_i^t\) is the local neighbor density beyond threshold, computed adaptively based on agent’s tree depth, and \(\mathcal{P}_i^t\) is the cumulative penalty from rule violations during action execution.

\paragraph*{\textbf{Relay Reward}}

Relay agents are responsible for preserving the integrity of the communication backbone and facilitating reliable data flow toward the root. Their reward function is structured to encourage stable connectivity while discouraging excessive burden on any individual relay. In addition to sustaining link quality for downstream agents, a key objective is to promote balanced load distribution across the relay network, thereby avoiding local congestion or bottlenecks that could compromise system scalability and responsiveness. The local reward for an relay agent $i$ at time $t$ is defined as:

\begin{equation}
	r^{\text{relay}}_i(t) = w_4 \cdot \mathbb{I}_{\{\text{goal held}\}} - w_5 \cdot \delta_i^t - \mathcal{P}_i^t,
\end{equation}

\noindent where \(\mathbb{I}_{\{\text{goal held}\}}\) indicates that agent \(i\) currently serves as a communication bridge for a discovered goal, and \(\delta_i^t\) is the deviation of current load from network-wide average.

\paragraph*{\textbf{Penalty Term \(\mathcal{P}_i^t\)}}

The penalty term $\mathcal{P}_i^t$ captures violations arising during agent execution. These include out-of-bound movements beyond the map, position conflicts due to attempted occupation of an already occupied location, and connectivity-related violations. The latter covers disconnection from the parent node (critical for maintaining upward links), loss of child connections (especially for relay agents), and structural breakage caused by movement. Each violation type contributes to $\mathcal{P}_i^t$ with role-specific severity weights.

\paragraph{Global Reward for Critic Training}  
To train the centralized critic \(Q_\phi(s^t,u^t)\), we compute a global reward \(R^t\) that balances task success and deployment cost:  

\begin{equation}
	\label{eq:global_reward}
	R^t =
	\underbrace{W_1 \cdot \mathcal{S}_{comm}^t + W_2 \cdot \mathcal{E}^t + W_3 \cdot \mathcal{X}^t}_{\text{Mission Performance}}
	-\;
	\underbrace{W_4 \cdot N^t + W_5 \cdot \mathcal{F}^t}_{\text{Cost and Penalty}} .
\end{equation}

Here, $\mathcal{S}_{comm}^t \!\in [0,1]$ evaluates communication quality by combining normalized end-to-end delay and bottleneck bandwidth over all target user-to-base paths (averaged across target users). $\mathcal{E}^t$ is the number of discovered and connected target users, while $\mathcal{X}^t$ measures exploration efficiency as the fraction of newly explored cells. $N^t$ penalizes redundant agent usage, and $\mathcal{F}^t$ captures failures such as path disconnection or tree collapse. Minor shaping terms are added to encourage smooth coverage and avoid stagnation. The reward is shared globally among agents and used only for critic training.

\paragraph{Reward Integration for Learning}

Each agent’s reward used for actor update combines its local role-specific reward and the global team reward:

\begin{equation}
	r_i^t = \lambda_{\text{local}} \cdot r_i^{\text{local}}(t) + \lambda_{\text{global}} \cdot R^t.
\end{equation}

This composite reward is stored in the agent’s replay buffer and guides its policy optimization. Meanwhile, the centralized critic uses only the global reward \(R^t\) to evaluate joint actions, enabling learning aligned with overall system goals. Also, penalties \(\mathcal{P}_i^t\) are embedded in the local rewards to discourage invalid behaviors without centralized intervention, supporting robust decentralized execution.

\subsection{On-Demand Agent Deployment Module}

To enable adaptive expansion with minimal coordination overhead, we treat deployment as an \emph{action-triggered process} tightly integrated with reinforcement learning, rather than a heuristic post-processing step. In this process, only explorer agents are allowed to initiate deployment requests, which are propagated upward along the logical deployment tree $\mathcal{T}_c(t)$, rooted at the base station, together with aggregated local statistics to support informed and resource-efficient decisions:

\begin{equation}
	\mathcal{R}_i = \left\{ \text{State}(a_j) \,\middle|\, a_j \in \text{Ancestors}(a_i) \cup \{a_i\} \right\},
\end{equation}

\noindent where each agent state includes load \(L_j\), exploration efficiency \(\epsilon_j\), and depth \(d_j\). The exploration productivity \(\epsilon_j\) is computed as:
\begin{equation}
	\epsilon_j = \frac{|\mathcal{G}_j^{\mathrm{explored}}(t_{\mathrm{recent}}) - \mathcal{G}_j^{\mathrm{explored}}(t_{\mathrm{past}})|}{t_{\mathrm{recent}} - t_{\mathrm{past}}},
\end{equation}

\noindent measuring the average number of new cells explored by agent \(a_j\) over a recent time window. This reflects the agent’s ability to expand the known environment frontier.

The base station validates the deployment request using two local criteria of the requesting agent \(a_i\). Specifically, it computes:

\[
\eta_i = \frac{|\mathcal{G}_i^{\mathrm{explored}}|}{|\mathcal{G}_i^{\mathrm{accessible}}|}, \quad
\delta_i = \frac{|\mathcal{N}_i|}{|\mathcal{G}_i^{\mathrm{neighborhood}}|},
\]

\noindent representing the local exploration ratio and neighborhood density. A request is approved only if \(\eta_i \geq \eta_{\min}\) and \(\delta_i \leq \delta_{\max}\), ensuring agents are deployed to underexplored, non-congested regions.

If approved, a parent node is selected from \(\mathcal{R}_i\) to anchor the new agent. The deployment score for each candidate \(a_j\) is given by:

\begin{equation}
	S_j = \lambda_1 (1 - \hat{L}_j) + \lambda_2 \hat{\epsilon}_j + \lambda_3 \hat{d}_j,
\end{equation}

\noindent where \(\hat{L}_j = L_j / L_{\max}\), \(\hat{\epsilon}_j = \epsilon_j / \epsilon_{\max}\), and \(\hat{d}_j = d_j / d_{\max}\) are normalized statistics. The weights \(\lambda_1, \lambda_2, \lambda_3 \in \mathbb{R}^+\) balance load, exploration, and structural growth.

\begin{figure*}[!t]
	\centering
	\includegraphics[width=5in]{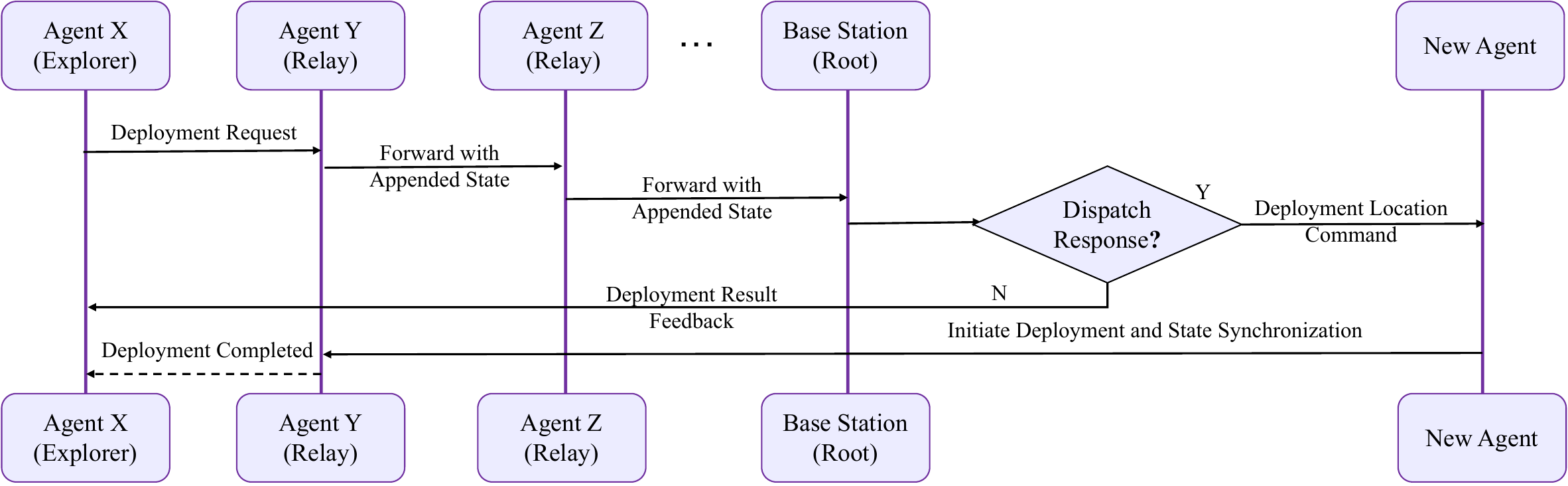}
	\caption{Illustration of the Dynamic Agent Dispatch Process.}
	\label{DP}
\end{figure*}

The system selects a parent node using a softmax-weighted score, prioritizing low-load, high-utility candidates while allowing adaptive variability. The new agent is then deployed near the selected parent and integrated into \(\mathcal{T}_c(t)\), enabling spatially aware, load-balanced network growth. Fig.~\ref{DP} illustrates this on-demand dispatch pipeline: an explorer triggers a request, statistics are aggregated along the deployment tree, and a new agent is dispatched to extend the network adaptively.

\section{A³ Network Architecture and Learning Framework}

This section details the core components of the proposed A³ network, including the decentralized deployment strategy for agents, the adaptive backhaul path selection process for efficient data routing, and the centralized training with decentralized execution paradigm enabled by GNN-based critics. The overall learning and deployment algorithm is summarized in Algorithm~1.

\subsection{Data Backhaul Path Selection and Routing Metrics}

In the proposed multi-agent system, agents are incrementally deployed to explore unknown environments while maintaining reliable backhaul connectivity to the base station. Operating under partial observability, each agent constructs and updates its path to the base using only local neighborhood information (Fig.~2(d)). 

Each agent maintains a locally constructed path to the base station, guided by three routing metrics: cumulative delay \(\mathcal{D}_i\), bottleneck available capacity \(\mathcal{C}_i\), and accumulated load \(\mathcal{L}_i\). Upon deployment, these are initialized as \(\mathcal{D}_i = \infty\), \(\mathcal{C}_i = 0\), \(\mathcal{L}_i = 0\) for all \(a_i \ne \mathrm{BS}\), and \(\mathcal{D}_{\mathrm{BS}} = 0\).

At each step, agent \(a_i\) evaluates neighbors \(a_j \in \mathcal{N}_i\) and computes candidate path metrics based on upstream values and link properties:
\begin{equation}
	\begin{split}
		D_{ij} &= \mathcal{D}_j + \left( \frac{d_{ij}(t)}{v} + \frac{L_i}{C_{ij}^{\mathrm{avail}}(t)} \right), \\
		C_{ij} &= \min(\mathcal{C}_j, C_{ij}^{\mathrm{avail}}(t)), \quad
		L_{ij} = \mathcal{L}_j + L_i,
	\end{split}
\end{equation}

\noindent where \(d_{ij}(t)\), \(v\), and \(L_i\) follow the definitions in Section~II. The routing metric uses the available link capacity \(C_{ij}^{\mathrm{avail}}(t)\) to account for bandwidth utilization.

Each candidate neighbor is assigned a weighted score:
\begin{equation}
	S_{ij} = w_D \cdot D_{ij} + \frac{w_C}{C_{ij}} + w_L \cdot L_{ij},
\end{equation}

\noindent where \(w_D\), \(w_C\), and \(w_L\) are tunable weights balancing delay, bandwidth, and load. The neighbor minimizing \(S_{ij}\), subject to feasibility constraints (e.g., sufficient available capacity and lower delay), is selected as the next hop, and metrics are updated accordingly.

Once a target user is detected, the agent forwards data hop-by-hop along its relay path, while intermediate nodes update load and link status in real time. Loop freedom is naturally ensured, as any cycle would only increase delay and thus be dominated. Instead of relying on RL actions, routing is maintained via distributed optimization, allowing the network to adapt to new deployments or changing conditions without global knowledge. This tree-consistent mechanism integrates smoothly with the overall learning and deployment framework in Algorithm~1.

\begin{algorithm}
	\caption{Training and Deployment of A³ Network via Actor-Critic with GNN-based Actor and Critic}
	\KwIn{Initial agent \( a_0 \), grid environment, actor params \( \{\theta_i\} \), critic params \( \phi \)}
	\KwOut{Trained decentralized actor policies \( \pi_{\theta_i} \)} 
	Initialize env., control tree \( \mathcal{T}_c(0) \leftarrow \{a_0\} \), comm. graph \( \mathcal{G}_c(0) \)\;
	\For{each episode}{
		Reset env. and agent states\; 
		\While{episode not terminated}{
			\For{each agent \( a_i \in \mathcal{A}(t) \)}{
				Observe \( o_i^t \), act \( u_i^t \sim \pi_{\theta_i}(\cdot \mid o_i^t, \mathcal{N}_i(t)) \)\;
				\If{\(a_i\) is explorer and requests deployment}{
					Send request upward along \( \mathcal{T}_c(t) \)\;
				}
			}
			Execute joint action, update env., update \( \mathcal{G}_c(t) \), dispatch new agents if approved\;
			\For{each agent \( a_i \)}{
				Propagate metrics (delay \( \mathcal{D}\), capacity \( C\), load \( L\)) upstream\;
				Select next-hop relay, update path \( \mathcal{P}_{i \rightarrow \text{BS}}(t) \)\;
				\If{target user discovered}{initiate hop-by-hop backhaul}\;
			}
			Compute global reward \( r^{\text{global}} \)\;
			\For{each agent \( a_i \)}{
				Local reward \( r_i^{\text{local}} \);\,
				Shaped reward \( r_i^{\text{shaped}} = \alpha r_i^{\text{local}} + (1-\alpha) r^{\text{global}} \);\,
				Store transition\;
			}
			\If{training}{
				Critic estimates advantage from \( r^{\text{global}} \);\,
				Update \( \theta_i \) with shaped reward and critic advantage;\,
				Update \( \phi \) via TD-error\;
			}
		}
	}
\end{algorithm}

\subsection{Training and Testing Process}

The proposed A³ system is trained under a CTDE paradigm. During training, each agent operates based solely on its local observation \( o_i^t \), while a centralized critic accesses the full communication graph \( \mathcal{G}_c(t) \) to estimate a global value function. This structure enables scalable policy learning without sacrificing decentralized autonomy during deployment.

The critic is implemented via a GNN, which encodes \(\mathcal{G}_c(t)\) into latent node embeddings and estimates action-value functions \( Q_\phi(\mathcal{G}_c(t), u) \) for joint actions \(u\). Actor parameters \( \theta_i \) are updated via policy gradients that maximize the expected Q-values using local rewards \( r_i^{\text{local}} \), while the critic parameters \( \phi \) are optimized with global rewards \( r^{\text{global}} \) reflecting system-level objectives such as connectivity, load balancing, and exploration.

To stabilize training and mitigate the issue of sparse global rewards, each actor is updated with a weighted combination of local and global rewards. This shaping accelerates convergence and avoids oscillations. Meanwhile, the critic is trained solely with global rewards, ensuring consistency with system-level objectives. Thus, the reward design balances local responsiveness with global alignment, preventing potential mismatch.

Once training converges, actor policies \( \{\pi_{\theta_i}\} \) are fixed, and during testing, agents operate fully decentralized using only local observations and internal states, without the critic. This ensures scalability and robustness, with performance measured by target user discovery, backhaul success, communication stability, and agent utilization.

\section{Numerical Results}

\begin{table}[htbp]
	\centering
	\caption{System Configuration Parameters in the A³ Network}
	\begin{tabular}{@{}p{0.5cm} p{5.5cm} p{2cm}@{}}
		\toprule
		\textbf{Para.} & \textbf{Description} & \textbf{Default (Unit)} \\
		\midrule
		\multicolumn{3}{@{}l}{\textbf{Environment and Agent Settings}} \\
		\midrule
		$\mathcal{E}$ & 2D discrete environment grid & $10 \times 10$ (cells) \\
		$\mathbf{p}_\mathrm{BS}$ & Base station position & $(0, 0)$ (--) \\
		$N_{\max}$ & Maximum number of deployable agents & 12 (--) \\
		$r_c$ & Communication radius of each agent & 3 (cells) \\
		$P_{\mathrm{tx}}$ & Transmission power $P_t$ & 1.0 (W) \\
		$G_t$ & Transmitter antenna gain $G_t$ & 1.0 (--) \\
		$G_r$ & Receiver antenna gain $G_r$ & 1.0 (--) \\
		$\lambda$ & Signal wavelength $\lambda_c$ & 0.125 (m) \\
		$B$ & Communication bandwidth & $10M$ (Hz) \\
		$N_0$ & Channel noise power $\sigma^2$ & $10^{-9}$ (W) \\
		$L_{\max}$ & Maximum communication load per agent & 50 (unit/task) \\
		\midrule
		\multicolumn{3}{@{}l}{\textbf{Target User-Related Settings}} \\
		\midrule
		$|\mathcal{T}|_{\max}$ & Maximum number of target users randomly placed in the environment & 5 (--) \\
		$L_i$ & Target user workload, randomly sampled in $[5, 15]$ & $\leq 15$ (unit/task) \\
		$C_i^{\min}$ & Minimum capacity requirement, randomly sampled in $[1, 3]$ & (Mbps) \\
		$D_i^{\max}$ & Maximum delay constraint, randomly sampled in $[30, 60]$ & (ms) \\
		\midrule
		\multicolumn{3}{@{}l}{\textbf{Global Settings}} \\
		\midrule
		$\theta_{\text{explore}}$ & Exploration sufficiency threshold for deployment requests & 0.75 (ratio) \\
		$\theta_{\text{load}}$ & Load threshold for deployment eligibility (as fraction of capacity) & 0.8 (ratio) \\
		$N_{\max}^{\text{deploy}}$ & Maximum number of deployment requests per agent & 3 (times) \\
		\midrule
		\multicolumn{3}{@{}l}{\textbf{Actor-Critic Learning Settings}} \\
		\midrule
		$\gamma$ & Discount factor for future returns & 0.95 (--) \\
		$\alpha_\pi$ & Actor learning rate & $1 \times 10^{-3}$ (--/step) \\
		$\alpha_V$ & Critic learning rate & $1 \times 10^{-3}$ (--/step) \\
		$|\mathcal{B}|$ & Batch size for actor-critic updates & 128 (samples) \\
		$|\mathcal{M}|$ & Replay buffer capacity & 800 (transitions) \\
		$\epsilon_0$ & Initial exploration noise & 0.1 (std) \\
		$\epsilon_{\min}$ & Minimum exploration noise & 0.01 (std) \\
		$\eta_\epsilon$ & Exploration decay rate & 0.995 (per step) \\
		$w_{\text{global}}$ & Global reward weight & 0.5 (--) \\
		$w_{\text{local}}$ & Local reward weight & 0.5 (--) \\
		$\mathcal{A}_{\text{move}}$ & Movement dimension (up/down/left/right/stay) & 5 (--) \\
		$\mathcal{A}_{\text{deploy}}$ & Deployment action dimension (binary: deploy or not) & 2 (--) \\
		\bottomrule
	\end{tabular}
	\label{tab:system_config}
\end{table}

\subsection{Simulation Settings} 

We evaluate the proposed A³ Network in a custom $10 \times 10$ grid, where agents collaboratively explore and serve randomly distributed target users under communication constraints. Each grid cell represents the minimal sensing and communication unit, allowing clear behavior observation without added complexity. Simulation settings are summarized in Table~\ref{tab:system_config}. Up to 12 agents may be deployed per episode; once this limit is reached, no further new agents are introduced. Wireless links follow a free-space path loss model, with per-link capacity computed via Shannon’s theorem. Wireless link capacity is calculated using Shannon’s theorem. Given that propagation delays are much smaller than transmission delays and our map is an abstract representation without precise physical distances, we neglect propagation delay. Additionally, since relay agents keep link distances short, this simplification is justified.

Given decentralized and autonomous decision-making, it is crucial to handle execution conflicts. While many multi-agent systems assume parallel and conflict-free execution, this often overlooks real-world issues such as collisions or connectivity loss. To address this, we adopt simultaneous decision-making with sequential execution and runtime validity checks. Invalid actions—those causing collisions or disconnections—are rejected, and the agent retains its previous state. This practical relaxation enables explicit conflict detection and resolution during simulation.

\subsection{Numerical Results and Analysis}

Our simulation-based evaluation demonstrates three key advantages of the proposed \textbf{A³ Network}: (1) stable coordination during training, (2) superior target user search capability in scaling environments, and (3) communication-aware resource efficiency.

\subsubsection{Comparative Baselines} 

To evaluate our adaptive framework, we compare it with two non-adaptive baselines under identical communication constraints. Learning-based or MARL-based methods are not included here, as most existing approaches are designed for different assumptions, such as full observability or known target user locations, which do not align with our decentralized and partially observable setting. Many of these methods rely on static optimized deployments that mainly serve as theoretical upper bounds rather than practical baselines. For fair comparison, we therefore select one such static upper-bound scheme and one non-adaptive coverage strategy without target user awareness, representing a reasonable performance range for our method.

\textbf{1) Static Deployment (Greedy-GA, Heuristic Upper Bound).} 
This baseline adopts a static, offline deployment strategy that combines greedy initialization with a multi-objective genetic algorithm (GA). The greedy phase generates feasible initial solutions with full target user coverage and connectivity, while the GA performs iterative refinement through mutation (agent addition, removal, relocation) and topology-aware crossover. A Pareto-based selection mechanism is used to balance deployment cost, end-to-end delay, and path capacity. Infeasible solutions are discarded, and the final Pareto front provides an approximate upper bound under full environment knowledge, serving as a benchmark for evaluating adaptive or learning-based approaches.

\textbf{2) Greedy Max-Coverage.} 
Agents are incrementally placed to maximize spatial coverage under communication connectivity constraints, without considering any information about target user demands. The placement process seeks to optimize a coverage metric while ensuring that each agent can maintain communication links within a fixed range. Although not explicitly designed for search or target user connection, this strategy provides a meaningful reference for evaluating how well a non-adaptive, exploration-only strategy performs in unknown environments.

Both baselines follow the same communication range constraint. One assumes full prior knowledge and optimal deployment, while the other explores blindly without awareness of target users. Outperforming both, our method proves superior in balancing adaptability, coordination, and mission success under decentralized uncertainty.

\begin{figure}[!t]
	\centering
	\includegraphics[width=3in]{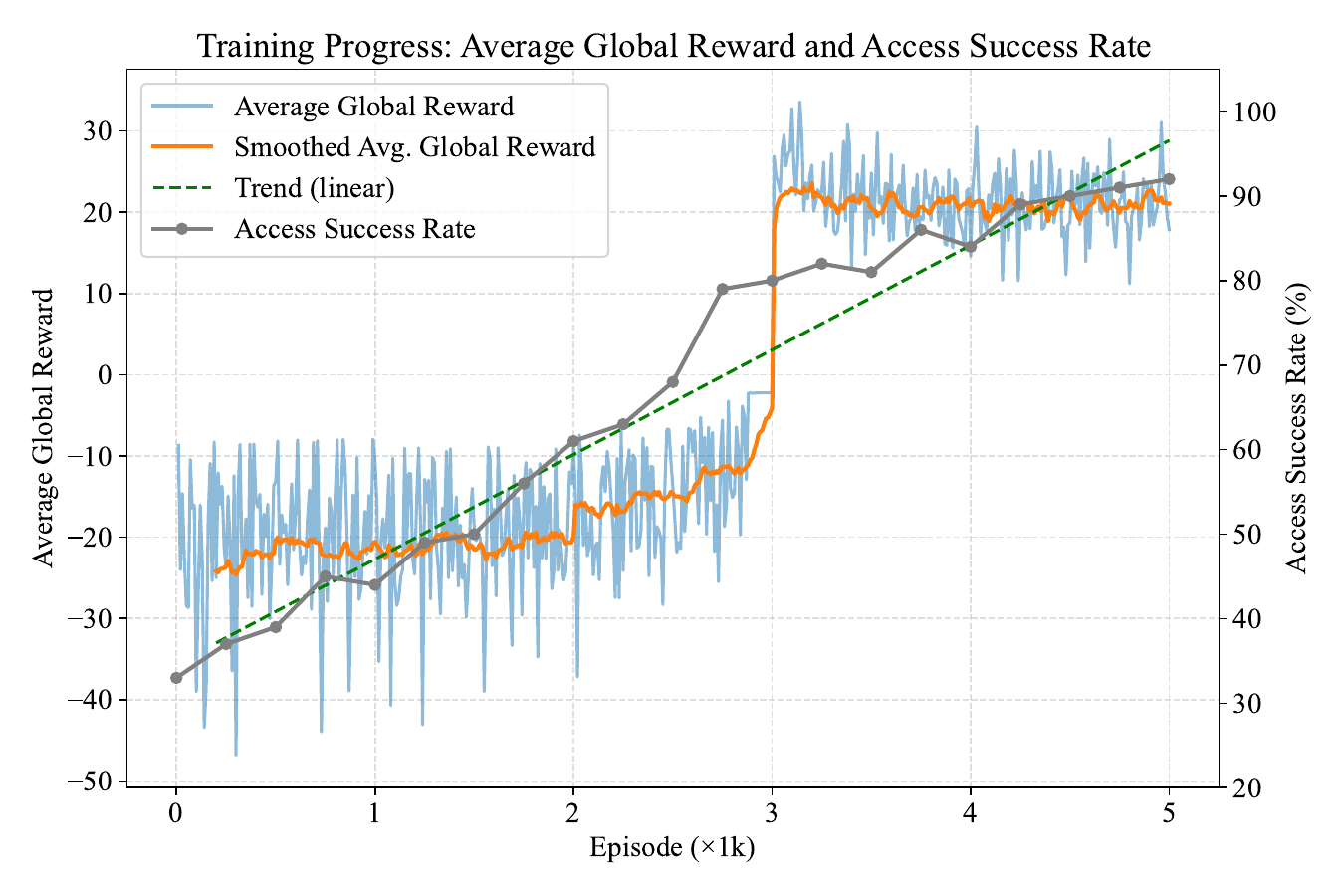}
	\caption{Evolution of average reward and access success rate during training episodes.}
	\label{R11}
\end{figure}

\subsubsection{Emergent Learning Behavior and Policy Convergence}

Fig.~\ref{R11} reports the evolution of the average reward during training. In the initial phase (0–5k steps), exploratory behavior dominates, leading to large fluctuations without a clear upward trend. This is expected since agents have not yet learned effective coordination. As training progresses (5k–10k steps), cooperative behavior gradually emerges, reflected by a rising reward trend and reduced variance. Beyond 10k steps, the system reaches a relatively stable phase with only minor fluctuations. While fluctuations remain due to stochastic exploration and partial observability, the joint increase of both reward and task success indicates that the learned policy stabilizes around a reliable operating regime rather than collapsing to local minima.

In addition to the reward trajectory, Fig.~\ref{R11} also presents the access success rate, defined as the probability of discovering all target users and establishing a communication link within the allowed agent budget. As training advances, this success rate increases steadily alongside the average reward, eventually exceeding 90\%. This consistency across two distinct metrics provides stronger evidence of convergence than reward alone. To avoid over-interpretation, we note that “convergence” here does not imply strict optimality but rather practical stability sufficient for deployment. This notion of convergence is consistent with common practice in MARL literature, where stability and reproducibility are often emphasized over strict optimality under partial observability.

\subsubsection{Target User Search Performance in Scaling Environments}

\begin{figure*}[!t]
	\centering
	\includegraphics[width=5in]{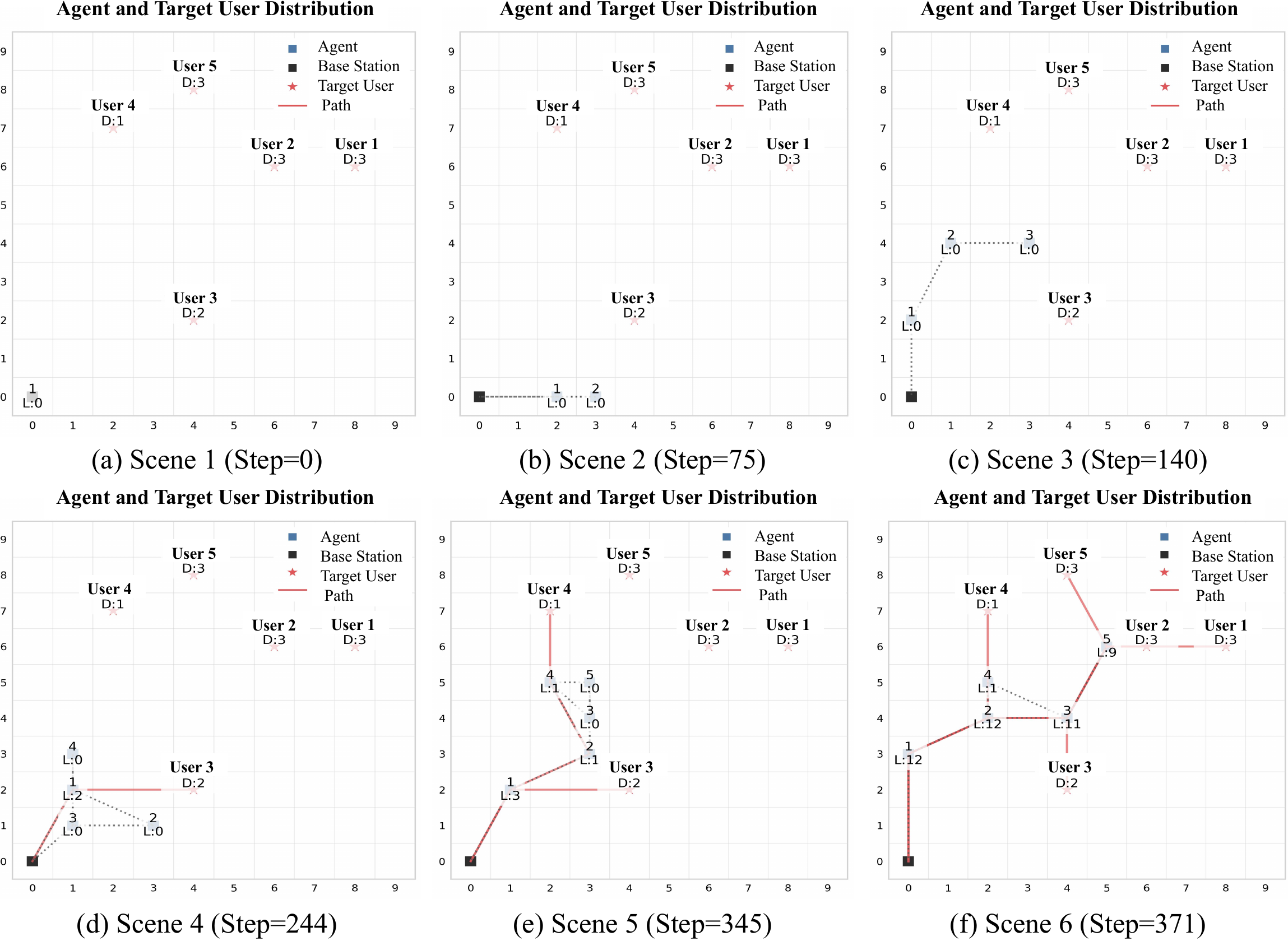}
	\caption{Illustration of the A³ Network exploring an unknown environment.}
	\label{R21}
\end{figure*}

\begin{figure*}[!t]
	\centering
	\includegraphics[width=5in]{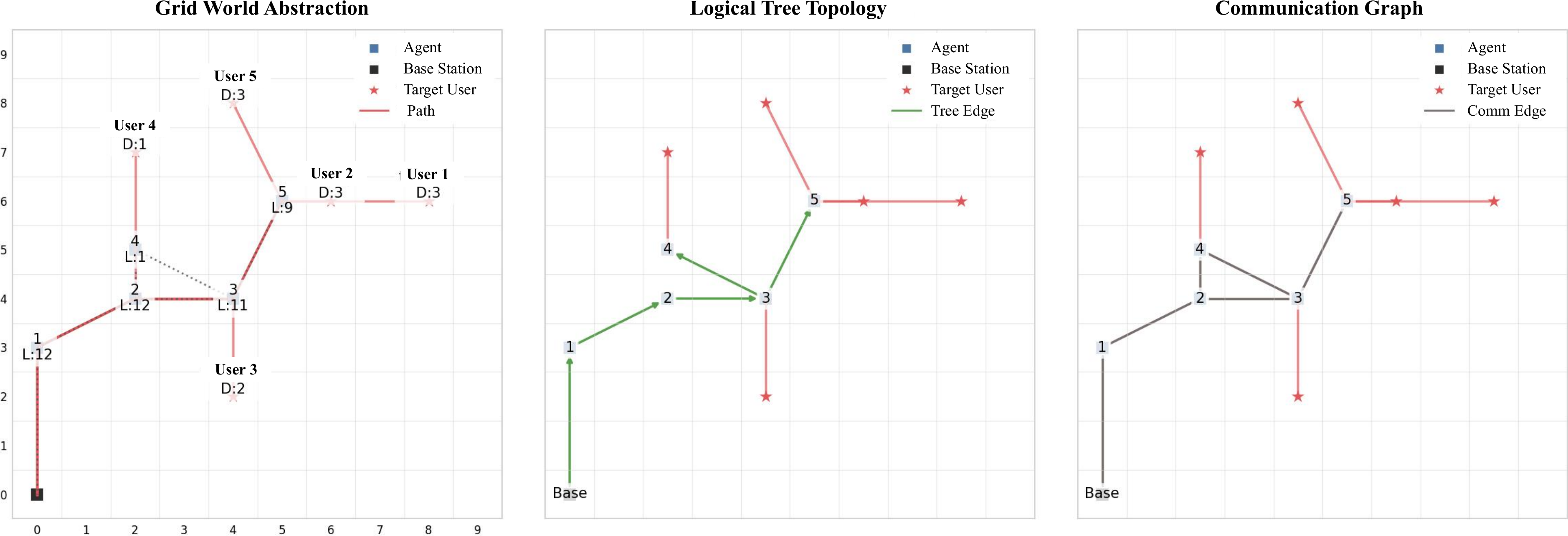}
	\caption{Adaptive Exploration and Communication-Aware Deployment by A³ Network.}
	\label{R22}
\end{figure*}

Fig.~\ref{R21} illustrates the process by which the A³ Network explores an unknown environment, searches for target users, and establishes communication links. Scene~1 depicts the initial stage of exploration, where the first agent is deployed from the base station, with no prior knowledge of the target users' locations. Scenes~2 through~5 show the progressive deployment of additional agents as the system expands its search area in pursuit of the target users. Scene~6 demonstrates the successful discovery of all target users and the establishment of a complete communication chain linking the target users back to the base station.

To further clarify the resulting topology, Fig.~\ref{R22} presents the corresponding relationship between the logical spanning tree and the physical communication graph formed at the moment all target users are connected. In the logical tree, the direction of each arrow represents the parent-child relationship between agents. To clearly represent service demand, each target user's load requirement is annotated as $D$, while the cumulative load handled by each agent is denoted as $L$. As target users are discovered and integrated into the network, their service loads are incrementally propagated along the relay path and distributed among the intermediate agents.

\begin{figure}[!t]
	\centering
	\includegraphics[width=3.5in]{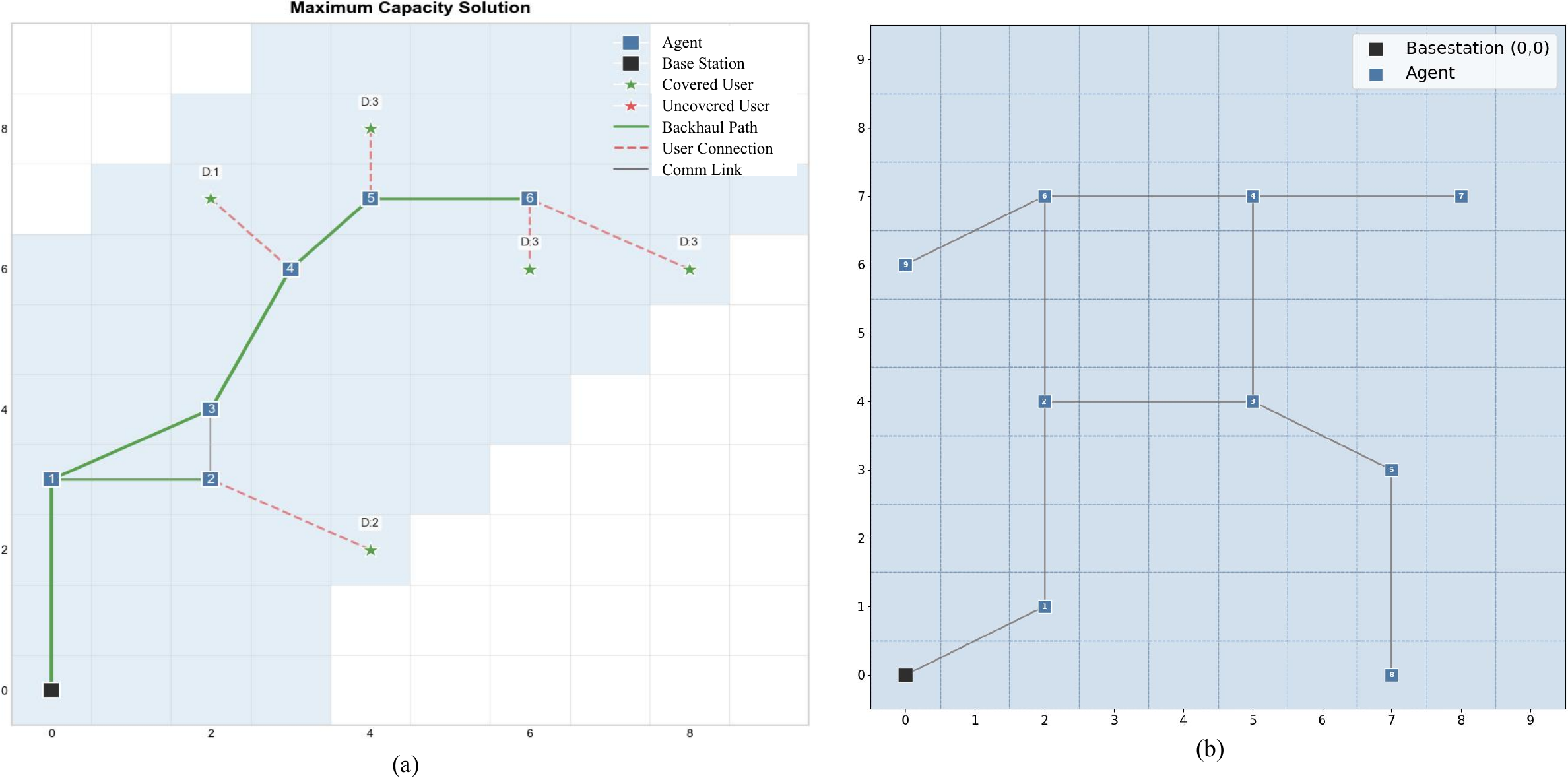}
	\caption{Baseline Deployment Strategies under Connectivity-Constrained Target User Service.}
	\label{R23}
\end{figure}

To benchmark against alternatives, Fig.~\ref{R23} compares A³ with two non-adaptive baselines: (a) a static offline optimization assuming full prior target user knowledge (serving as an upper bound rather than a deployable strategy), and (b) a greedy max-coverage heuristic ignoring target user's demand. While these baselines do not represent learning-based methods, they reflect the two extremes of practical deployment—omniscient static planning versus blind coverage-first placement. We acknowledge that alternative RL baselines exist, but to our knowledge they either assume centralized information or do not support on-demand deployment, making a direct comparison not meaningful in this setting. Nevertheless, our chosen baselines allow us to quantify both the theoretical limit (static with full knowledge) and a natural heuristic (coverage-only), thereby bracketing the achievable performance space.

\begin{figure}[!t]
	\centering
	\includegraphics[width=3.6in]{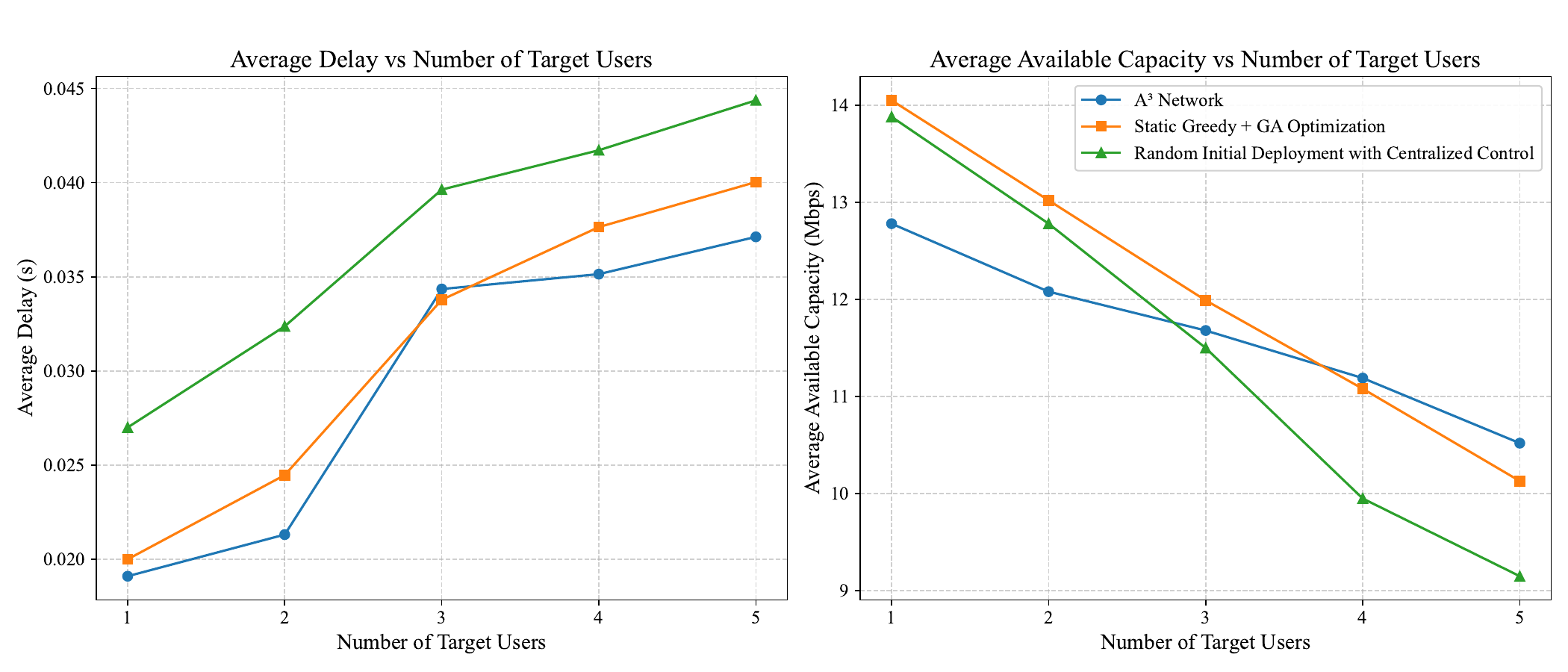}
	\caption{Average Delay and Average Capacity vs. Number of Target Users for Different Deployment Strategies.}
	\label{R3}
\end{figure}

\begin{figure}[!t]
	\centering
	\includegraphics[width=3.5in]{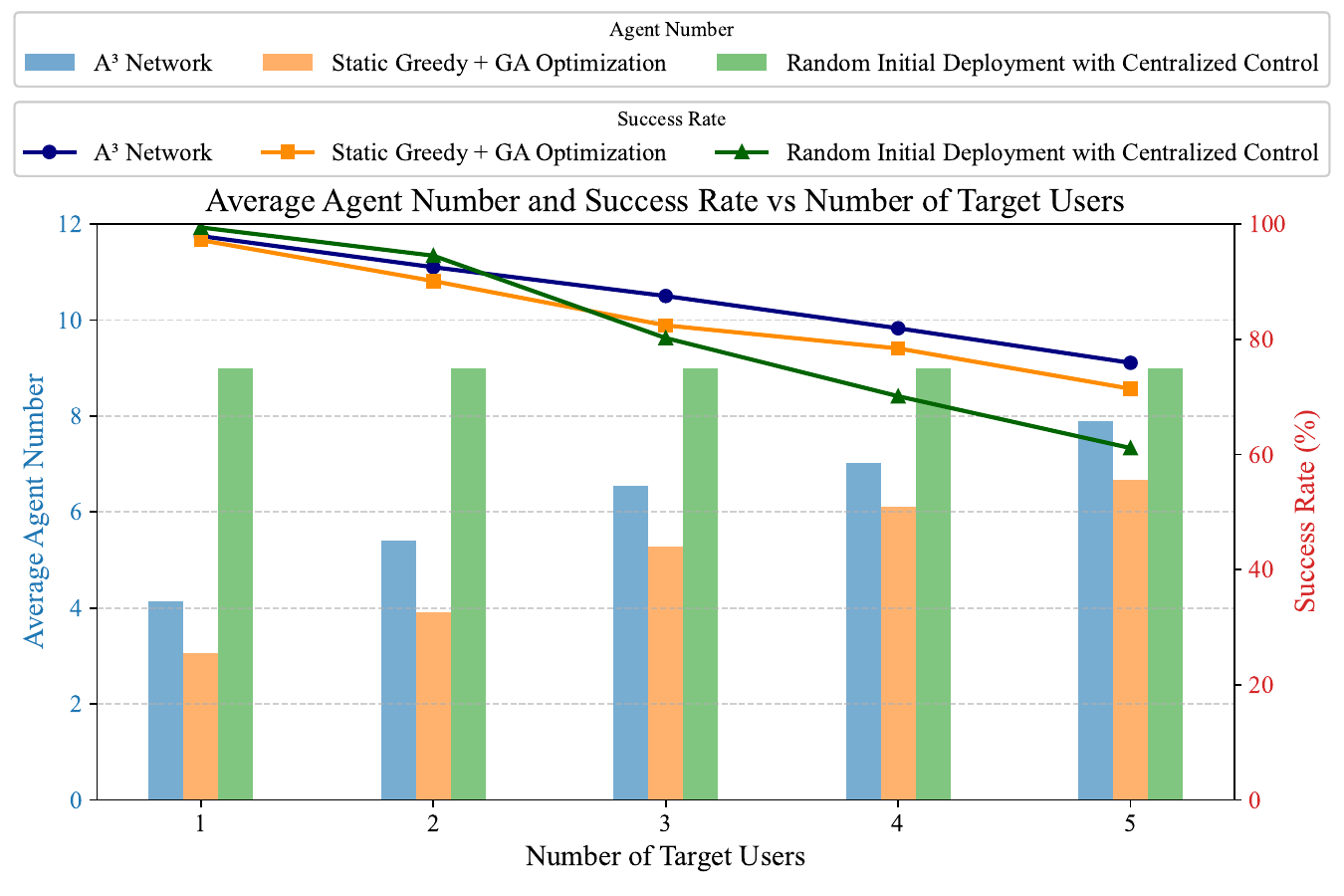}
	\caption{Average Agent Number and Success Rate vs. Number of Target Users for Different Deployment Strategies.}
	\label{R4}
\end{figure}

\subsubsection{Communication-Aware Performance Benchmarking}

We further evaluate communication-aware performance in a $10\times10$ grid with varying numbers of target users. As the number of target users increases, agents are reused across multiple relay paths, leading to higher load per node. To capture this, we measure both average end-to-end delay and effective per-link capacity.

Since \textbf{Greedy Max-Coverage} completely disregards both target user distribution and communication constraints, we introduce a modified baseline, \textit{Random Initial Deployment with Centralized Control}, to assess performance under a coverage-first yet communication-unaware policy. In this scheme, agents are randomly placed at initialization (without incremental dispatch) and then follow a centralized, rule-based policy designed to preserve connectivity while maximizing coverage. This setup allows us to isolate the impact of a coverage-first but communication-agnostic approach on delay, capacity, and success rate.

As shown in Fig.~\ref{R3} (Average Delay and Average Capacity), the \textit{A³ network} consistently delivers superior communication performance. It maintains the lowest delay across varying target user counts and sustains capacity with only minor degradation. In contrast, \textit{Static Greedy + GA} performs well under light load but exhibits a sharper decline in capacity and slightly higher delay as target user density increases. \textit{Random Initial Deployment} shows the weakest performance, with persistently higher delay and the steepest capacity drop. These results underscore the advantages of A³’s dynamic, communication-aware design: (i) it adaptively balances load and preserves channel quality under growing demand, (ii) static optimization methods provide limited short-term gains but degrade rapidly, and (iii) coverage-first heuristics significantly compromise end-to-end communication performance.

Fig.~\ref{R4} (Average Agent Number and Success Rate) highlights the trade-off between resource usage and communication reliability. \textit{Random Initial Deployment} uses the largest agent pool and achieves high initial success rates at low target user density due to wide, simultaneous coverage. However, this advantage is short-lived: the strategy emphasizes spatial coverage without considering communication needs, resulting in poor alignment between agent positions and target user connectivity. As target user density grows, the lack of coordination and adaptability leads to sharp performance degradation, with agents unable to maintain reliable links despite their large numbers.

In contrast, the \textit{A³ network} dynamically adjusts agent deployment based on both spatial distribution and communication constraints, ensuring that coverage is functionally useful. By jointly considering agent load, connectivity, and evolving target user's demand, A³ maintains the highest overall success rates while using resources efficiently. Compared to this, \textit{Static Greedy + GA} performs reasonably under low-density conditions but lacks the adaptability to sustain performance as network demands grow.

\subsubsection{On-Demand Deployment Efficiency and Scalability} 

As shown in Table~\ref{tab:scaling}, the \textit{Greedy Max-Coverage} strategy consistently requires the largest agent count across all grid sizes, as it aims for complete spatial coverage regardless of actual target user or communication needs. 

The \textit{Static Greedy+GA Optimization} scheme achieves the smallest agent count in most cases; however, this efficiency comes at a cost. It relies on full prior knowledge of target user locations and fixes agent positions before deployment, making it unable to adapt if target users move or if any agent fails.

In contrast, the proposed A³ Network combines low agent usage with strong adaptability. For example, in the $10 \times 10$ environment, it requires an average of 6.89 agents, which is fewer than \textit{Greedy Max-Coverage} (9) and only slightly more than \textit{Static Greedy + GA Optimization} (5.92). In larger environments ($12 \times 12$ and $14 \times 14$), the A³ Network deploys 10.16 and 14.22 agents respectively, significantly fewer than \textit{Greedy Max-Coverage} at 14 and 18, and close to the static scheme’s 8.28 and 12.13 agents. Importantly, A³ achieves this performance without access to prior target user's information, dynamically adjusting its deployment to match real-time communication demands.

\begin{table}[htbp]
	\centering
	\caption{Comparison of Agent Numbers and Network Performance Across Different Environment Grid Sizes}
	\begin{tabular}{c c c c c c}
		\hline
		\textbf{Metric / Grid Size} & \textbf{6$\times$6} & \textbf{8$\times$8} & \textbf{10$\times$10} & \textbf{12$\times$12} & \textbf{14$\times$14} \\
		\hline
		Greedy Max-Coverage & 4.00 & 6.00 & 9.00 & 14.00 & 18.00 \\
		A³ Network                    & 2.35 & 3.91 & 6.89 & 10.16 & 14.22 \\
		Static Greedy+GA Optimization & 2.71 & 3.84 & 5.92 & 8.28  & 12.13 \\
		\hline
	\end{tabular}
	\label{tab:scaling}
\end{table}

These results underscore A³ Network’s ability to minimize redundancy while ensuring coverage and connectivity. Its adaptive, on-demand dispatch enables efficient use of resources, making it well-suited for large-scale or dynamically changing environments where both exploration and communication performance are important.

\section{Conclusion}

This work introduced the A³ Network, a communication-aware MARL framework that addresses exploration, target user access, and backhaul formation under partial observability in uncertain environments. By decoupling physical connectivity from logical coordination, and by training policies with CTDE over dynamic communication graphs, A³ enables scalable cooperation and on-demand deployment without requiring prior target user's information. Evaluations in scaling grid environments further show that A³ maintains reliable connectivity and efficient agent usage while sustaining communication performance. Rather than pursuing strict optimality, the framework achieves practical convergence and stable coordination, demonstrating robustness for deployment in dynamic and information-limited scenarios. 

Future work will extend A³ to more complex environments with physical obstacles and line-of-sight constraints. The decentralized control and on-demand deployment mechanisms are naturally suited to such conditions, as local decision-making and adaptive relay dispatch allow the network to circumvent blockages and cover irregular boundaries. Furthermore, scaling from 2D to 3D spatial domains will support scenarios such as aerial–ground coordination and integrated air–space–terrestrial networks, advancing toward the vision of intelligent, infrastructure-free 6G communication systems.

\vfill

\end{document}